\pdfoutput=1
\documentclass[sigconf]{acmart}

\AtBeginDocument{%
  }

\copyrightyear{2026}
\acmYear{2026}
\setcopyright{cc}
\setcctype{by}
\acmConference[UIST '26]{The 39th Annual ACM Symposium on User Interface Software and Technology}{November 02--05, 2026}{Detroit, MI, USA}
\acmBooktitle{The 39th Annual ACM Symposium on User Interface Software and Technology (UIST '26), November 02--05, 2026, Detroit, MI, USA}
\acmDOI{10.1145/3830398.3830692}
\acmISBN{979-8-4007-2856-3/2026/11}

\ifdefined\HIGHLIGHTREVISIONS
  \newcommand{\rev}[1]{{\color{blue!70!black}#1}}
\else
  \newcommand{\rev}[1]{#1}
\fi

\definecolor{dgInk}{HTML}{B84A6E}
\newcommand{\dlab}[1]{\textcolor{dgInk}{\textbf{#1}}}

\begin{document}

\title[The Many Lives of Personal Data]{Supporting The Many Lives of Personal Data with Rebite: LLM-Powered Goal-Directed Framing in Food Journaling}%
\author{Weijun Li}
\email{weijunl3@uci.edu}
\orcid{0000-0001-6068-5185}
\affiliation{%
  \institution{University of California, Irvine}
  \city{Irvine}
  \state{California}
  \country{USA}
}

\author{Daniel A. Epstein}
\email{epstein@ics.uci.edu}
\orcid{0000-0002-2657-6345}
\affiliation{%
  \institution{University of California, Irvine}
  \city{Irvine}
  \state{California}
  \country{USA}
}

\begin{abstract}
People's health and tracking goals frequently change, but most personal informatics systems struggle to adapt, leading people to abandon their data and start over. We propose \emph{goal-directed framing}, \rev{an approach that repositions goals within personal informatics systems}. Instead of fixing the meaning of data at capture time, the approach frames the collected data through the current goal and reframes it whenever the goal changes. We realize this in \textit{Rebite}, a photo-based food journaling system that uses LLMs to read unstructured meal photos and produce goal-directed feedback. In a one-week deployment with 21 participants managing multiple dietary goals, we find that goal-directed framing shaped how participants engaged with their goals. Translating a goal into metrics helped them see what it meant in practice, confirming existing priorities, surfacing what they overlooked, and revealing where the metrics fell short. When goals changed, seeing past meals reframed under the new goal exposed overlaps and conflicts, prompting participants to negotiate trade-offs and refine priorities. We discuss how goal-directed framing both supports and complicates reflection as goals change, and offer design implications for personal informatics systems to support evolving goals.

\end{abstract}


\begin{teaserfigure}
  \centering
  \includegraphics[width=\textwidth]{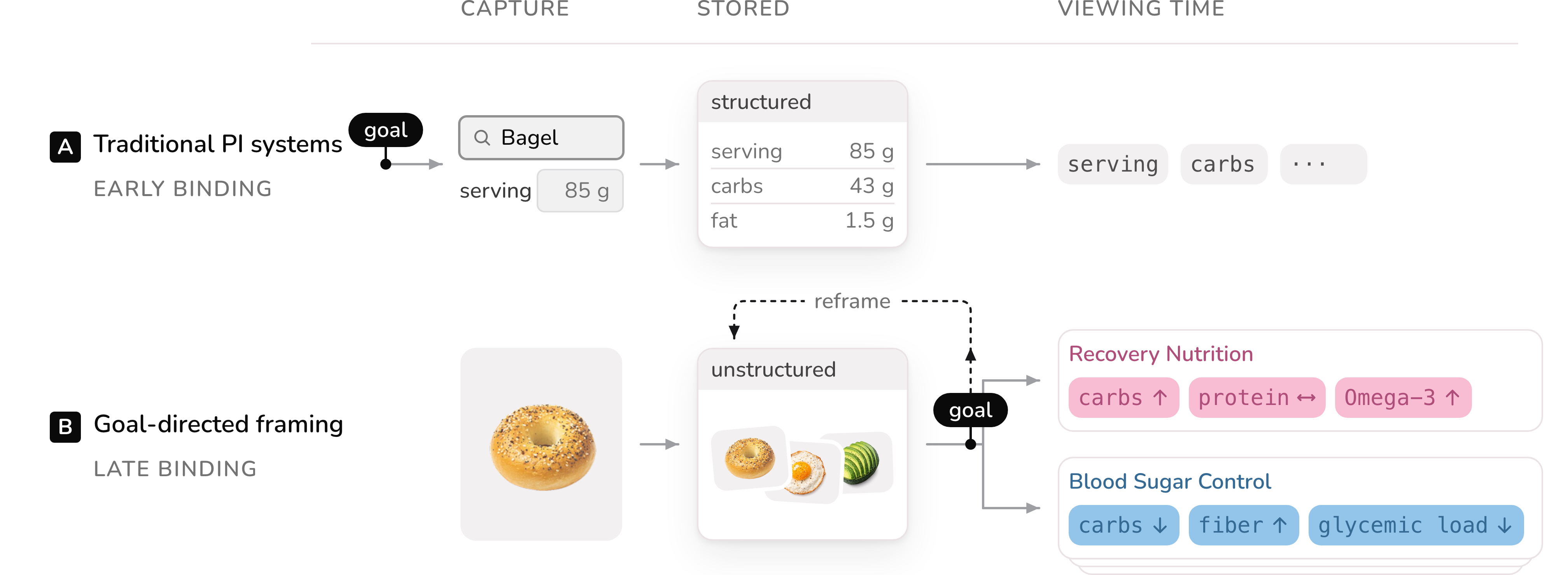}
  \caption{(\textbf{A})~Traditional Personal Informatics systems fix the metrics when data are logged, so a goal adopted later relies on the same structured data. (\textbf{B})~Our approach of Goal-directed framing keeps data as unstructured and applies the goal at viewing, so one record can be interpreted and reframed to support more than one goal.}
  \label{fig:paradigm}
\end{teaserfigure}

\maketitle
\section{Introduction}
\label{sec:intro}

Personal informatics research has long argued that aligning the data people collect with their goals supports reflection and action~\cite{li2010stage,mapping2020epstein,schroeder2020goal}. Yet, goals change. People revise, refine, expand, or abandon their goals as routines, health needs, and priorities evolve~\cite{mapping2020epstein,catrin2022Ttransitions,sefidgar2024migraine,epstein2015lived}. Goals may be translated from qualitative aspirations into concrete, quantitative targets~\cite{niess2018fitness}, and may also multiply, evolve, and shift in priority over time~\cite{elena2022goal}. When the system's metrics no longer match the user's evolved goal, people often suspend tracking or switch to new tools, leaving past data behind~\cite{epstein2015lived,lu2021socially,catrin2022Ttransitions,clawson2015no}. How to build PI systems that adapt when goals change remains largely unexplored.

Why do current PI systems fail to support goal change? Goals evolve in at least two ways. They may gradually become clearer as users learn what matters~\cite{niess2018fitness,elena2022goal}, or shift as priorities change~\cite{catrin2022Ttransitions,ekhtiar2025goal}. In both cases, the interpretation of existing data should also change, including which metrics are relevant, at what level of detail, and how to present them for reflection. Yet, most current systems do not fully adapt. They either hard-code which metrics to present~\cite{moore2022bacon} or allow users to configure tracking only prospectively~\cite{kim2017omni,sefidgar2024migraine}. As a result, the meaning of collected data remains tied to decisions made at logging time. When goals change, users often cannot reinterpret their existing data through the new goal because the information needed for that interpretation was never preserved or made available.

We propose \textbf{goal-directed framing}, \rev{an approach that repositions goals within personal informatics systems where goals shape interpretation at the time of viewing data.} The approach preserves rich, unstructured data (such as photos) and, when the user states a goal, collaboratively translates it into relevant metrics and structures feedback through that goal's lens. When the goal changes, the system \emph{reframes}, by reinterpreting all past data under the new goal, so the same data acquires new meaning without re-logging. We use large language models (LLMs) to realize this, as they can reason over unstructured inputs and generate goal-directed feedback~\cite{shin2025fitting}.

We realize goal-directed framing in \textbf{Rebite}, a photo-based food journaling system (Figure~\ref{fig:pipeline-walkthrough}). Food journaling is a useful testbed for this approach because people pursue a range of dietary goals that change over time~\cite{ha2025life,ekhtiar2025goal}, from weight loss~\cite{cordeiro2015barriers} and managing chronic conditions like diabetes~\cite{mitchell2021r2c} to identifying food triggers~\cite{Karkar2017TummyTrials} and eating more mindfully~\cite{epstein2016crumbs}. Meal photos carry rich nutritional detail across many dimensions (e.g., calories, macronutrients), and interpreting it under a specific goal can require expertise that most users do not have~\cite{luo2019food,cordeiro2015photo}. Structured support that connects meal photos to the user's current goal is therefore important for meaningful reflection.

To understand how people experience goal-directed framing in practice, we deployed Rebite with 21 participants managing multiple dietary goals in a one-week study with a mid-study goal switch. We find that the goal-to-metric mapping helps participants make a vague goal concrete, confirming what they already care about, surfacing what they overlook, and revealing where the metric space falls short. When goals change, reframing past meals under the new goal exposes overlaps and conflicts between goals, prompting participants to negotiate trade-offs and reconsider what their goals mean in practice. We discuss how goal-to-metric mapping enables reframing, how reframing brings both benefits and risks, and what it means for personal data to stay meaningful as goals change. We make three contributions.

\begin{itemize}
  \item \rev{\emph{An approach for PI systems with evolving goals.} We propose goal-directed framing, which repositions the goal within personal informatics systems} by having the system take on the role of translating the current goal into relevant metrics, framing preserved data through that goal's lens, and reframing it when the goal changes, enabling personal data to ``live many lives.''
  \item \emph{Design and implementation of Rebite.} A photo-based food journaling system that realizes goal-directed framing, letting the same collected data be reflected upon under evolving goals without new data entry.
  \item \emph{Empirical evidence from a deployment study ($N$\,=\,21).} We illustrate how reframing past data under new goals can make cross-goal relationships visible, including overlaps that validated past efforts, conflicts that prompted trade-off negotiation, and breakdowns that revealed when users preferred to start fresh instead of revisiting.
\end{itemize}

\section{Background and Related Work}

To motivate our approach, we first review \emph{goals in personal informatics} and examine \emph{when interpretations become fixed in existing systems}. We then discuss \emph{LLMs for personal data} as creating new opportunities for flexible interpretation, and finally situate our work within the context of \emph{food journaling}.

\subsection{Goals in Personal Informatics}

Goals are central to what and how people track. Prior work shows that goals guide what people record and how they reflect on their data~\cite{li2010stage,schroeder2020goal,li2021reflection}. In lived practice, however, goals evolve with everyday routines, life transitions, and health needs~\cite{mapping2020epstein,catrin2022Ttransitions,sefidgar2024migraine,epstein2015lived}, and people revise, expand, or abandon them accordingly.

People progressively sharpen goals from abstract needs to quantifiable targets~\cite{niess2018fitness}, select among multiple problems, simplify them into specific goals, and iteratively readjust~\cite{elena2022goal}, and refine goals through reflection~\cite{lee2015personlized}. The direction and pace of change are shaped by personal, contextual, and social factors~\cite{ekhtiar2025goal,saksono2024socio}. A recent scoping review argues that PI systems should be designed to support goal adjustment over time~\cite{ekhtiar2023goal}.

Prior systems have pursued alignment between goals and data through flexible configuration of tracking fields~\cite{kim2017omni}, supporting selection of apps that match tracking needs~\cite{lee2022app}, and presenting data in ways that reflect stated goals~\cite{sefidgar2024migraine,gouveia2023watch}. Specifically, Schroeder et al.~\cite{schroeder2020goal} proposed goal-directed self-tracking as a framework in which designs centered around an individual's goals support tracking exactly and only the data that individual needs. MigraineTracker~\cite{sefidgar2024migraine} provides empirical evidence for this approach, aligning data collection and presentation with clinical goals to support reflection and decision making. However, none of these systems reinterpret previously collected data when goals evolve. A growing body of work has called for PI systems that support goal adjustment~\cite{ekhtiar2023goal,ekhtiar2025goal,epstein2015lived,catrin2022Ttransitions}, but reinterpreting past data under a new goal has so far been done by hand where it has been done at all, and how it shapes people's engagement with their goals has not been studied. \rev{We address this gap by framing already-collected data through the goals a person already holds and moves between, complementing prior work on forming and refining goals~\cite{niess2018fitness,lee2015personlized,schroeder2020goal,ekhtiar2023goal}.}

\subsection{When Does Interpretation Get Locked In?}
\label{sec:binding-time}

Current PI systems bind interpretation early, in one of two ways, and as a result struggle to reinterpret past data when goals change. Many systems hard-code a single frame at design time. Calorie-focused food apps~\cite{tsai2007usability} foreground energy budgets, and photo-based nutritional analysis~\cite{noronha2011Platemate} extracts a fixed metric set when data enters. Moore et al.~\cite{moore2022bacon} call this the ``personal informatics analysis gap'', where tools ``baking-in'' analysis workflows, leaving users unable to ask questions the designers did not anticipate. Other systems give users control over what to track, but apply it only going forward. OmniTrack~\cite{kim2017omni} lets users compose custom fields, MigraineTracker~\cite{sefidgar2024migraine} lets patients select and adjust tracking routines, yet reconfiguration reaches only future data, and previously logged entries are not re-assessed. When a goal changes and the collected data no longer supports that goal, people often abandon the tool or start over~\cite{epstein2015lived,epstein2016abandon}. A growing line of food journaling work preserves richer, open-ended data to support reflection~\cite{lucas2025foodtalk,luo2021foodscrap}, but does not reinterpret it when goals change.

A nascent body of work points the other way, suggesting that reinterpretation is both desired and possible. Ng et al.~\cite{ng2022bounded} found that people bound periods of disruption in their data, such as a pregnancy, and interpret data inside and outside those boundaries differently, and Jung et al.~\cite{jung2025databox} showed that letting users manipulate their own tracked values can trigger new reflection. Design work has begun exploring open-ended tracking that resists fixing meaning at design time~\cite{toebosch2025triadic,barkercanler2024chromatize}.

\rev{The goal has long been seen as central to what and how people track~\cite{li2010stage,schroeder2020goal}, and prior systems let it shape what is collected and how it is shown. What no system does is re-apply it to data already in hand. The goal is bound once, prospectively, and never turned back on the record.} We call this missing capability \emph{goal-directed framing}, in which the system translates a goal into relevant metrics and frames collected data through that goal's lens, then \emph{reframes} it whenever the goal changes, so past data stays relevant instead of locked in its original frame.

\rev{This capability has overlap with strategies in interactive visualization, where user actions can transform views on data. For example, dynamic queries that update a visualization when the user adjusts a parameter~\cite{Ahlberg1992DynamicQueries}, and reactive visualization grammars that propagate a signal change through a dataflow graph to every dependent mark~\cite{Satyanarayan2014ReactiveVega}. In this way, the goal acts as a parameter, yielding different interpretations of an existing dataset. However, the parallel falls short in that visualization typically assumes the dataset is already captured, and is being interactively analyzed. In personal informatics, since goals emerge and change and future goals are hard to anticipate, it is instead unclear at collection time what to record and which dimensions will later matter. People also revisit and reanalyze their data in different ways than traditional visualization. PI practice favors smaller, more frequent reflection situated in everyday life, not a sit-down session with a dashboard~\cite{blascheck2021glanceable}. Supporting this kind of everyday reflection as goals change is thus a challenge PI needs to explore, and one our work takes up.}

\subsection{LLMs and Personal Informatics}

Recent work shows that LLMs can enhance personal informatics along two dimensions. For interpretability, LLMs have turned raw metrics into natural-language narratives to support reflection~\cite{stromel2024fitnees}, elicited experiential detail like emotions to complement numeric tracking~\cite{kim2024mindfuldiary}, and paired narrative summaries with visual timelines to help experts question and validate AI inferences~\cite{li2024vital}. For personalization, LLMs have translated goals into actionable IF--THEN plans~\cite{shin2025fitting}, generated context-sensitive journaling prompts from behavioral sensing data~\cite{nepal2024mindsape}, grounded activity coaching in wearable data~\cite{jorke2024supporting}, and let users choose among generated prompts and request summaries on demand~\cite{song2025self}.

These systems show that LLMs can personalize feedback and generate interpretive narratives around tracked data. Yet LLMs alone are not sufficient. Chopra et al.~\cite{Chopra2025llms} found that people querying generative AI about their health data had to keep scaffolding and refining their prompts to get goal-relevant answers, and abandoned queries when answers stayed too generic. Goal alignment therefore belongs in the system, not in the user's prompting. Prior work also suggests how to structure the output. Sensecape~\cite{suh2023sensecape} shows that organizing LLM output into levels of abstraction helps users navigate complex information, and Graphologue~\cite{jiang2023graphologue} shows that turning responses into interactive diagrams makes them more interpretable than plain text. This informs how Rebite presents goal-directed feedback in layers.

\subsection{Food Journaling}
Food journaling is a widely studied domain in personal informatics~\cite{mapping2020epstein}, and the focus of our implementation of goal-directed framing. People use food journaling to pursue a wide range of goals, including weight loss~\cite{cordeiro2015barriers}, managing chronic diseases~\cite{chuang2019photo,mitchell2021r2c}, and identifying intolerances~\cite{Karkar2017TummyTrials}. The practice also serves qualitative purposes, such as reflecting on the sensory and contextual aspects of meals~\cite{zhang2020thought} and building awareness through lightweight, mindful engagement~\cite{epstein2016crumbs}, and social purposes such as self-presentation~\cite{chung2017social}, social learning~\cite{lu2021socially}, and family support~\cite{lukoff2018tablechat}.

Researchers have explored multiple input modalities for food journaling. Calorie logs have been the traditional approach, leveraging database lookups of text descriptions~\cite{tsai2007usability} or barcodes. Photo-based journaling has been suggested as an alternative to balance entry difficulty~\cite{cordeiro2015photo}, with a long line of research seeking to derive nutrient estimates from meal images, whether by crowdsourcing~\cite{noronha2011Platemate} or by automated recognition~\cite{allegra2020review,tahir2021comprehensive}. Other strategies include conversational journaling~\cite{lucas2025foodtalk,luo2021foodscrap} and multi-device routines~\cite{lucas2023multimodal}. Each modality carries trade-offs. Text entry is tedious and often incomplete~\cite{cordeiro2015barriers,andrew2013simplifying}, image-based approaches reduce effort but struggle with portion estimation~\cite{allegra2020review,tahir2021comprehensive}, and semi-automated approaches raise privacy concerns~\cite{Xi2022semi}.

Food tracking goals frequently change. Life transitions such as moving or changing jobs reshape food routines and priorities~\cite{ha2025life}. What people track also differs across goals. An IBS patient follows symptom--food relationships~\cite{chuang2019photo,luo2019food}, someone managing type~2 diabetes needs personalized targets around carbohydrate intake~\cite{mitchell2021r2c}, and someone pursuing weight loss focuses on calories and macronutrients~\cite{cordeiro2015barriers}. Luo et al.~\cite{luo2019food} showed that different conditions require distinct tracking items and representations. Such customization, however, is typically set up once and does not support transitions between evolving goals. Cordeiro et al.~\cite{cordeiro2015photo} showed a decade ago that the value of photo capture varies with the journaler's goal, and argued that journals should support a variety of goals. What is missing is not the idea but a system that performs the reinterpretation, and evidence for what happens when it does.

\begin{figure*}[!t]
  \centering
  \includegraphics[width=\textwidth]{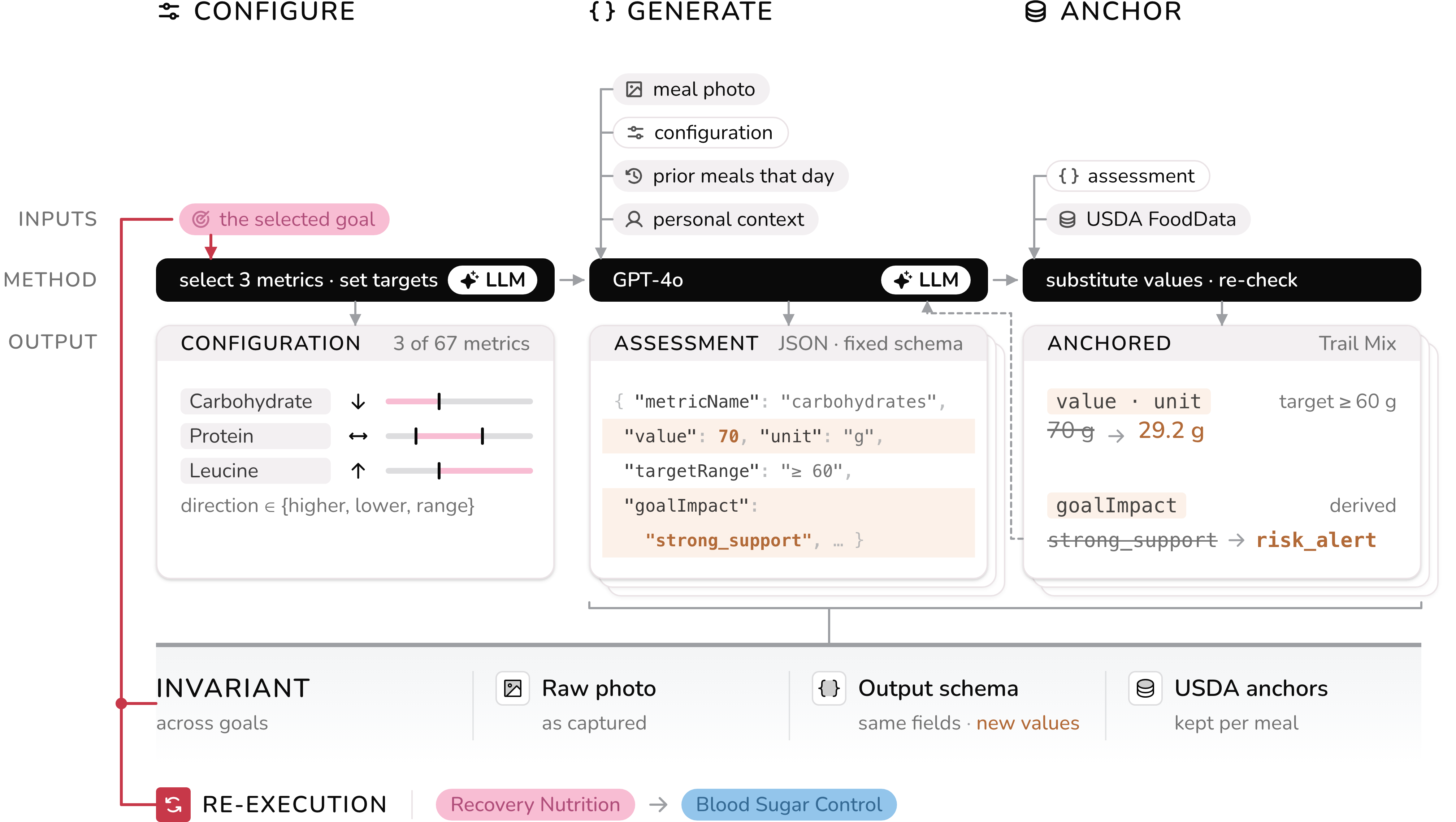}
  \caption{The interpretation pipeline that enables goal-directed framing. Based on a selected goal, the LLM creates a configuration of target metrics. When new data (meal photos) are logged, this configuration is passed alongside context to the LLM to generate a goal-specific assessment. This is then anchored against a food database. This pipeline can be re-executed on existing logs when a person's goal changes.}
  \label{fig:pipeline-walkthrough}
\end{figure*}

\section{Goal-Directed Framing: An Approach for Personal Informatics Systems with Evolving Goals}
\label{sec:paradigm}

As articulated in Section~\ref{sec:binding-time}, current PI systems lock interpretation because they bind meaning at design or setup time. To address this, we propose goal-directed framing. Instead of users conforming to the system's metrics, \emph{systems can serve the user's evolving intention}. Figure~\ref{fig:paradigm} contrasts this with current PI systems.

Realizing this model poses three design challenges, each derived from limitations of the strategies in Section~\ref{sec:binding-time}.

\noindent\dlab{D1.}~\emph{Translate goals into trackable configurations.} Goals range from vague aspirations to specific targets~\cite{niess2018fitness,Chopra2025llms}, so the system can propose metrics for a stated goal and let users refine them as their intent sharpens.

\smallskip\noindent\dlab{D2.}~\emph{Reframe past data when goals change.} Current systems interpret data once and apply new configurations only to future entries~\cite{kim2017omni,sefidgar2024migraine}, so a goal change has to send the system back over everything already stored. Reinterpretation only means something if the underlying facts hold still. If the recorded values themselves shift with the goal, a user cannot tell a new reading of the same meal from a new account of it.

\smallskip\noindent\dlab{D3.}~\emph{Bridge lightweight input to structured, goal-directed feedback.} Users should not have to specify metrics or label data. The system extracts goal-directed structure from unstructured input and presents it so that the goal connects to concrete assessments and evidence.

Because LLMs can reason over unstructured data and generate structured interpretations on demand, we propose an LLM-powered approach to address these challenges. The user logs unstructured inputs and states a goal, and the system does the mapping, interpretation and re-interpretation. Figure~\ref{fig:pipeline-walkthrough} shows this as we implement it in Rebite.

\rev{Goal-directed framing could extend to domains where rich raw data can be preserved and reinterpreted as goals change.} Exercise, sleep, and mental health are candidates, where sensor streams carry enough signal to be read through different goals. We instantiate and evaluate it in food journaling for the reasons given in Section~\ref{sec:intro}.

\section{The Rebite System}

\textit{Rebite} is our photo-based food journaling system that realizes goal-directed framing. Users log meals through photos and state a goal in everyday language. The system then translates the goal into trackable metrics~(\textbf{D1}), interprets each meal against those metrics and re-interprets all stored meals when goals change~(\textbf{D2}), and presents structured, four-layer feedback that moves from goal context to concrete food evidence~(\textbf{D3}).

\begin{figure*}[!t]
  \centering
  \includegraphics[width=\textwidth]{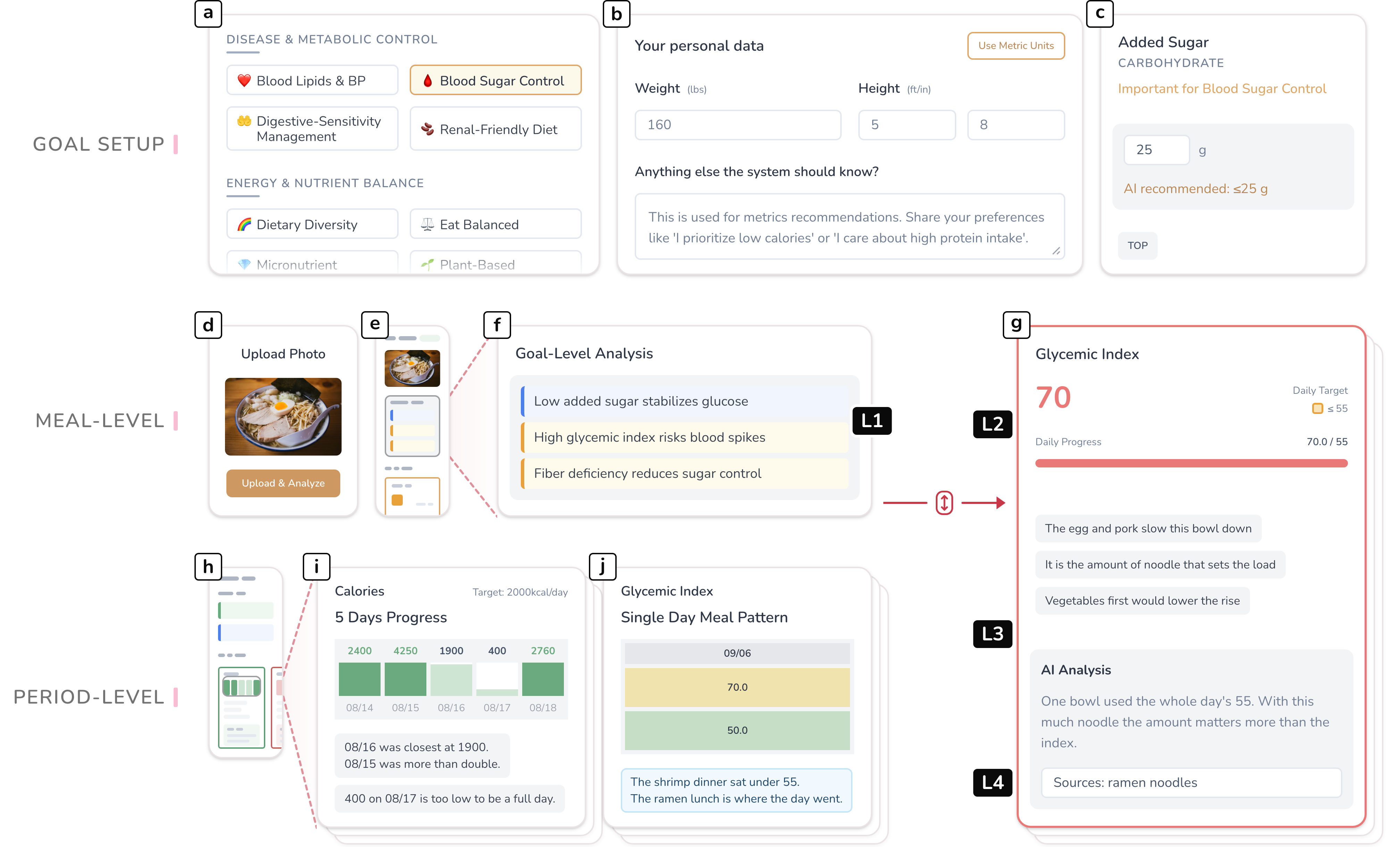}
  \caption{Goal-directed framing instantiated in Rebite. To set up their goal, a person (a)~selects from options and (b)~provides personal context, resulting in an (c)~AI-recommended configuration of metrics. When (d) new meal photos are logged, (e-f) they are summarized at the meal level with first layer L1 analysis of how they support the goal. (g) A person can scroll down for a metric card with L2 comparing progress against their target, L3 AI interpretation, and L4 specific food evidence. (h)~Data are summarized at a period-level, with visualizations for metrics which should be (i) increased or decreased or (j) should stay within a range. When goals change, each interpretation layer reframes existing data while interaction remains the same.}
  \label{fig:system}
\end{figure*}

\subsection{Goal Setup}
\label{sec:framework}

Users begin by selecting a goal in everyday language (e.g., ``lose weight,'' ``recover from training'') and optionally providing personal context such as height, weight, or health conditions (Figure~\ref{fig:system}a--b). The system then translates these into a concrete configuration of which metrics to track, in which direction (increase/decrease), and with what thresholds. Users can inspect the resulting configuration (Figure~\ref{fig:system}c) and adjust metrics or targets as their understanding deepens.

This translation is supported by a goal--metric framework that we compiled from published nutrition and public-health guidance (full mappings are provided in the supplementary material). The framework pairs 67 nutritional metrics with 19 goals, assigning each goal three priority metrics and several supporting ones, together with a direction and a starting threshold. For example, \textit{Lose Weight} prioritizes calories, protein, and fiber, whereas \textit{Blood Sugar Control} prioritizes carbohydrate dose, fiber, and glycemic load. The mappings draw on published nutrition guidelines and supporting evidence from the literature~\cite{USDA_HHS_DGA_2020_2025,ADA_Standards_2024,NASEM_DRI_EssentialGuide_2006,ellomartin2007energy,rolls2009energy}. Switching goals automatically produces a new configuration that reframes the same food records.

The framework provides a default configuration for each goal. Users can add, remove, or retarget metrics to better reflect their own priorities. They can also maintain multiple goals and switch among them, revisiting the same food records through different goal-directed framings.

\subsection{Goal-Directed Interpretation}
\label{sec:pipeline}

Rebite does not bind a meal to a single interpretation when it is logged. Instead, the same meal can be framed through the user’s \emph{current} goal configuration whenever feedback is requested.

\textbf{Meal logging.}
Users capture meals by uploading a photo (Figure~\ref{fig:system}d). After reviewing the system's analysis, users can refine the estimates by adding ingredient descriptions with quantities (e.g., ``100\,g pasta with 120\,g ground beef''), correcting misidentified items, or adjusting portions.\label{sec:stage2} Rebite then interprets each meal through three stages, where the output of each feeds the next (Figure~\ref{fig:pipeline-walkthrough}).

\textbf{Configure.}
The system sends the user's goal to the model together with the goal--metric framework (Section~\ref{sec:framework}). The model selects three priority metrics from the framework's library and personalizes each target's direction and threshold (the supplementary material gives an abridged version of all three pipeline prompts), producing a concrete configuration of which metrics to evaluate, in which direction, and with what thresholds. This configuration shapes all downstream interpretation.

\textbf{Generate.}
The system constructs a prompt from the meal photo, the configuration from `Configure', prior meals from the day, and the user's personal context. This prompt is sent to a VLM, GPT-4o in our implementation, which returns a structured JSON response following a fixed output schema. The response includes food identification, nutritional estimates for all tracked metrics, per-metric assessments relative to the goal, and goal-level tags. Because the configuration determines which metrics the LLM evaluates, the same photo produces different assessments under different goals.

\textbf{Anchor.}
The system looks up the foods the model identified in USDA FoodData Central and recomputes every tracked metric the database covers from its composition data and the model's estimated weights. What stays estimated is which foods are present and how much of each. Metrics the database does not carry, such as glycemic load and FODMAP content, stay model estimates, so goals resting on them keep the estimation error anchoring removes elsewhere. If the correction changes a metric's status relative to its target, the corrected values go back to the model, which re-generates that assessment.

When a user changes goals, `Configure' produces a new configuration, and `Generate' and `Anchor' re-execute on every stored meal. Both calls conform to a fixed output schema regardless of the active goal, so reframing requires no schema migration.

\subsection{Goal-Directed Feedback}
\label{sec:interface}

The interface is designed around the principle that \emph{the structure of feedback stays constant while its content reframes}. This gives users a stable visual language for interpreting goal-directed assessments, even as the underlying metrics, colors, and narratives change when goals change.

\textbf{Meal-level feedback.}
Each meal's feedback is organized as four layers that move from goal context to factual evidence (Figure~\ref{fig:system}e--g).
\textbf{Goal-level tags} color-code each nutritional dimension by its impact on the current goal, separating dimensions that support it from those that carry risk or ask to be monitored (Figure~\ref{fig:system}f, L1). When the goal changes, the same meal's tags change color, making reframing immediately visible.
\textbf{Metric cards} display the anchored value against the target (Figure~\ref{fig:system}g, L2), with a breakdown showing how this meal sits within the day. A status indicator shows how the metric performs relative to the goal.
\textbf{Goal-directed interpretations} explain what the metric value means for the user's specific goal (Figure~\ref{fig:system}g, L3).
\textbf{Food evidence} grounds each interpretation in specific ingredients from the meal (Figure~\ref{fig:system}g, L4).

\textbf{Period-level feedback.}
At the period level (e.g., a week), additive metrics (e.g., calories) use bar charts scaled against the metric's target (Figure~\ref{fig:system}i), while non-additive metrics (e.g., glycemic index) use heatmap grids (Figure~\ref{fig:system}j). Each claim lists the meals behind it and whether each one supports the goal, so a claim can be checked against the entries it came from. When goals change, the active metrics and visualization layout reframe~(\textbf{D3}).

\subsection{Implementation}
\label{sec:implementation}

Rebite extends the open-source \textit{Clean Slate} calorie tracker (Apache License 2.0)\footnote{\url{https://github.com/successible/cleanslate}} and is implemented as a web application using React and a GraphQL backend with a PostgreSQL database. The interpretation pipeline uses the OpenAI API to make calls to GPT-4o with JSON-constrained output. The pipeline can be adapted to other LLMs that support structured output. Before deploying Rebite with users, we ran a technical evaluation of how much reframing varies across goals, what factual anchoring contributes, and how much anchoring changes the values the model returns from a photo. The method and results are in the supplementary material.
\label{sec:implementation-validation}

\section{User Study}

\label{sec:user-study}

We conducted a mixed-method deployment study to understand how people experience goal-directed framing in practice, complemented by a pre/post survey for quantitative reference. The study protocol was reviewed and approved by our university's IRB, and all participants gave informed consent. We organize our investigation around two questions.

\noindent\dlab{RQ1.}~\emph{How do users make sense of goal-directed feedback?} When the system maps a vague goal into concrete metrics~(\textbf{D1}) and interprets meals through that lens~(\textbf{D3}), how do users experience the resulting feedback?

\smallskip\noindent\dlab{RQ2.}~\emph{What happens when goals change and past data is reframed?} When the system re-maps and re-interprets stored data under a new goal~(\textbf{D2}), what forms of reflection emerge, and where does the mechanism break down?

\subsection{Method}

Over one week, participants tracked meals under an initial goal for four days, then switched to a different goal mid-study. Because we were interested in how participants \textbf{perceived and made sense of} reframed feedback and not in measuring behavior change, we recruited people who already held multiple dietary goals and asked them to switch between existing priorities.
\begin{figure}[!t]
  \centering
  \includegraphics[width=\linewidth]{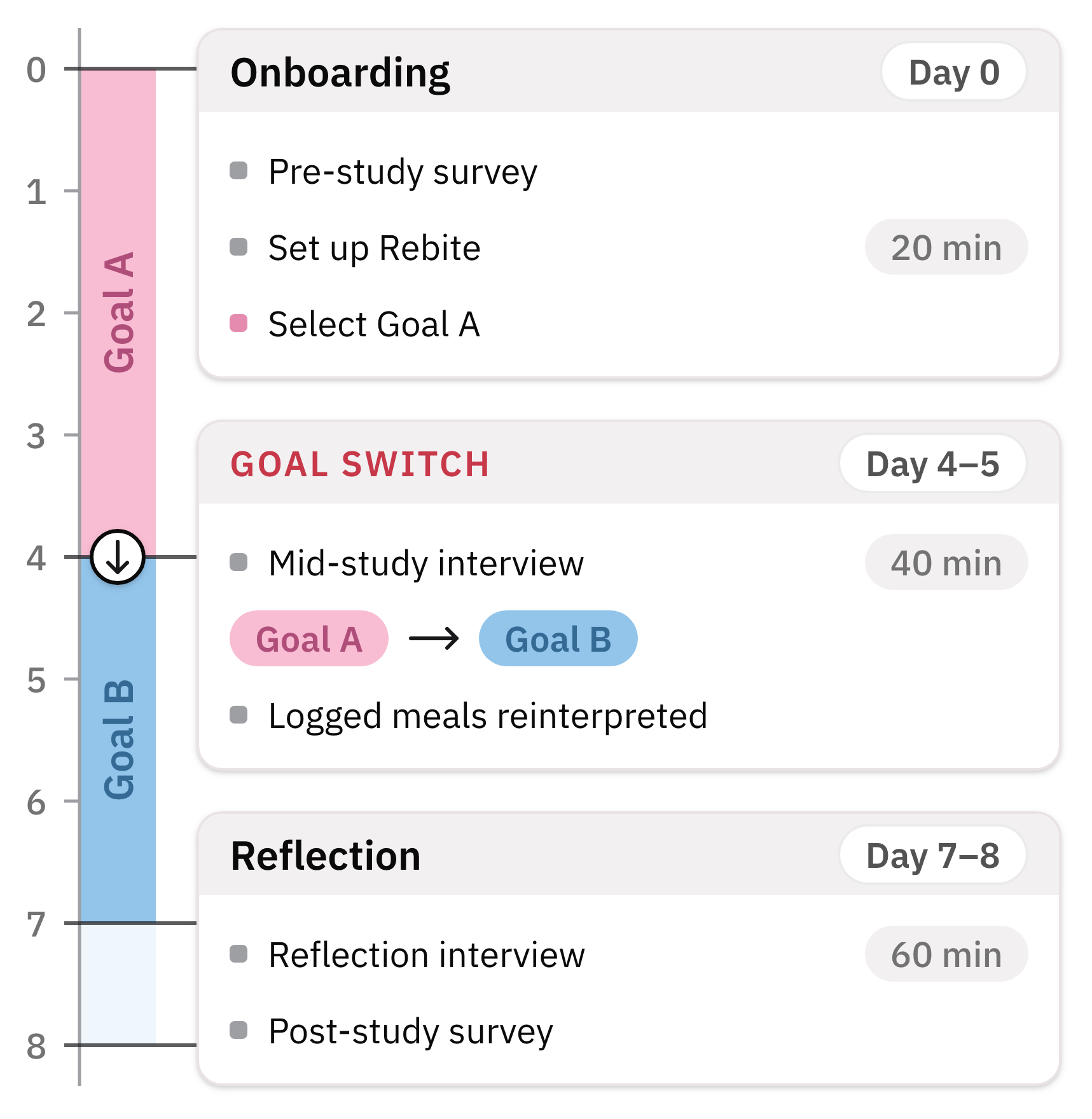}
  \caption{Study timeline ($N{=}21$). Participants logged meals under Goal~A, then switched to Goal~B following the mid-study interview.}
  \label{fig:study_timeline}
\end{figure}

Figure~\ref{fig:study_timeline} summarizes the schedule. Participants completed a pre-study survey about their prior tracking practices, then joined a 20-minute onboarding session where they set up Rebite on their phones and selected an initial goal (Goal~A). Over seven days, they logged meals with photos and reviewed feedback daily. After four days, participants joined a mid-study interview ($\sim$40 minutes) where they reviewed feedback under Goal~A, switched to a different goal (Goal~B), and discussed reinterpretations of their logged meals under the new goal. At the end of the study, participants completed an after-use survey and a 60-minute reflection interview. The full procedure is in the supplementary material.

\subsection{Participants}

We recruited 25 adults in the United States who owned a smartphone, had previously tracked meals for at least two weeks, and were currently pursuing multiple diet-related goals. The multiple-goal requirement kept the switch on existing priorities and off invented ones~\cite{epstein2015lived,niess2018fitness,ekhtiar2023goal}, and the prior-tracking requirement kept the study on goal-directed framing and off the novelty of food journaling. Because food tracking can be sensitive in the context of eating disorders, recruitment materials emphasized a range of dietary goals (e.g., muscle gain, anti-inflammatory eating, weight management). We recruited through flyers and diet-focused Discord communities.

Four participants withdrew during the study (citing usability issues and busyness), leaving $N{=}21$ (13F, 8M; ages 21--49). Each held two or three concurrent goals, 49 in all, spanning 14 of the framework's 19 goals. Weight loss ($n{=}14$) and muscle gain ($n{=}10$) were the most common, followed by anti-inflammatory eating and recovery nutrition ($n{=}4$ each), balanced eating and blood sugar control ($n{=}3$ each), and plant-based adequacy, reduced fat intake and glycogen reload ($n{=}2$ each). Gut health, mental clarity, blood lipids and blood pressure, micronutrient optimization, and digestive sensitivity were each held by one participant. Participants received an \$80 Amazon gift card on completion. Per-participant demographics, prior tracking experience, and goals are tabulated in the supplementary material.

\subsection{Analysis}

\subsubsection{Survey Analysis}
We administered a pre/post 11-item survey of 6 of the 9 Technology-Supported Reflection Inventory (TSRI) items~\cite{b2021tsri}, covering its Exploration and Insight dimensions, plus 5 goal-related items (Q7--Q11). We excluded its third dimension, Comparison, whose items concern comparison with other people. We reworded the TSRI items for the food domain and administered them on a 5-point scale, so we report them as TSRI-derived measures and not as the validated instrument. Pre and post used the same items, so the comparison is internal to that adaptation. Pre-study responses referred to participants' prior tools and post-study responses to Rebite. We compared scores with paired two-tailed $t$-tests, Cohen's $d_z$, and Holm--Bonferroni correction. The full instrument is in the supplementary material.

\subsubsection{Qualitative Analysis}
Transcripts from mid-study interviews, final interviews, and daily mini-surveys were coded in ATLAS.ti, combining deductive codes from the research questions and system features with inductive codes for emergent practices (e.g., goal conflicts surfaced by the system). Following reflexive thematic analysis~\cite{braun2006using,braun2019reflexive}, we emphasized depth over inter-coder reliability and kept analytic memos.

\subsubsection{Integration}
We used survey trends to locate where reflection scores changed and the qualitative data to explain how, cross-checking emergent themes against survey patterns to separate individual cases from broader trends.

\subsection{Study Limitations}
Although goals often evolve over months~\cite{ekhtiar2025goal}, we chose a one-week deployment to balance burden and engagement. \rev{The switch drew on goals participants already held, but the study scheduled it instead of letting it arise on its own, so our findings capture how a first reframing lands and not how reframing behaves across repeated, self-initiated changes or once novelty fades.} The survey compares Rebite against participants' recalled impressions of prior tools, which is subject to recall and novelty effects and is not a controlled baseline. \rev{We therefore read it as secondary evidence that reflection changed, and rest the findings themselves on the qualitative accounts.} Participants' goals clustered around common themes such as weight loss and muscle gain, leaving out rarer or culturally specific ones. Rebite relies on VLM interpretation, which our technical evaluation shows is only coarsely accurate from photos alone (supplementary material), and estimates remain imperfect for portion sizes and hidden ingredients. Our goal--metric framework emphasizes common health metrics and is an artifact we authored from published guidance, not a clinically validated instrument reviewed by a dietitian or clinician, so Rebite's feedback is not medical advice. Finally, we report what participants said about reframing, not how much they logged.

\section{Study Results}
\label{sec:study-results}

We organize findings around RQ1 and RQ2, using survey results as a quantitative reference and qualitative evidence to show how goal-directed framing shaped engagement with goals.

\subsection{Statistical Analysis of Pre--Post Survey Scores}\label{sec:stats}

\begin{figure}[!htbp]
  \centering
  \includegraphics[width=\linewidth]{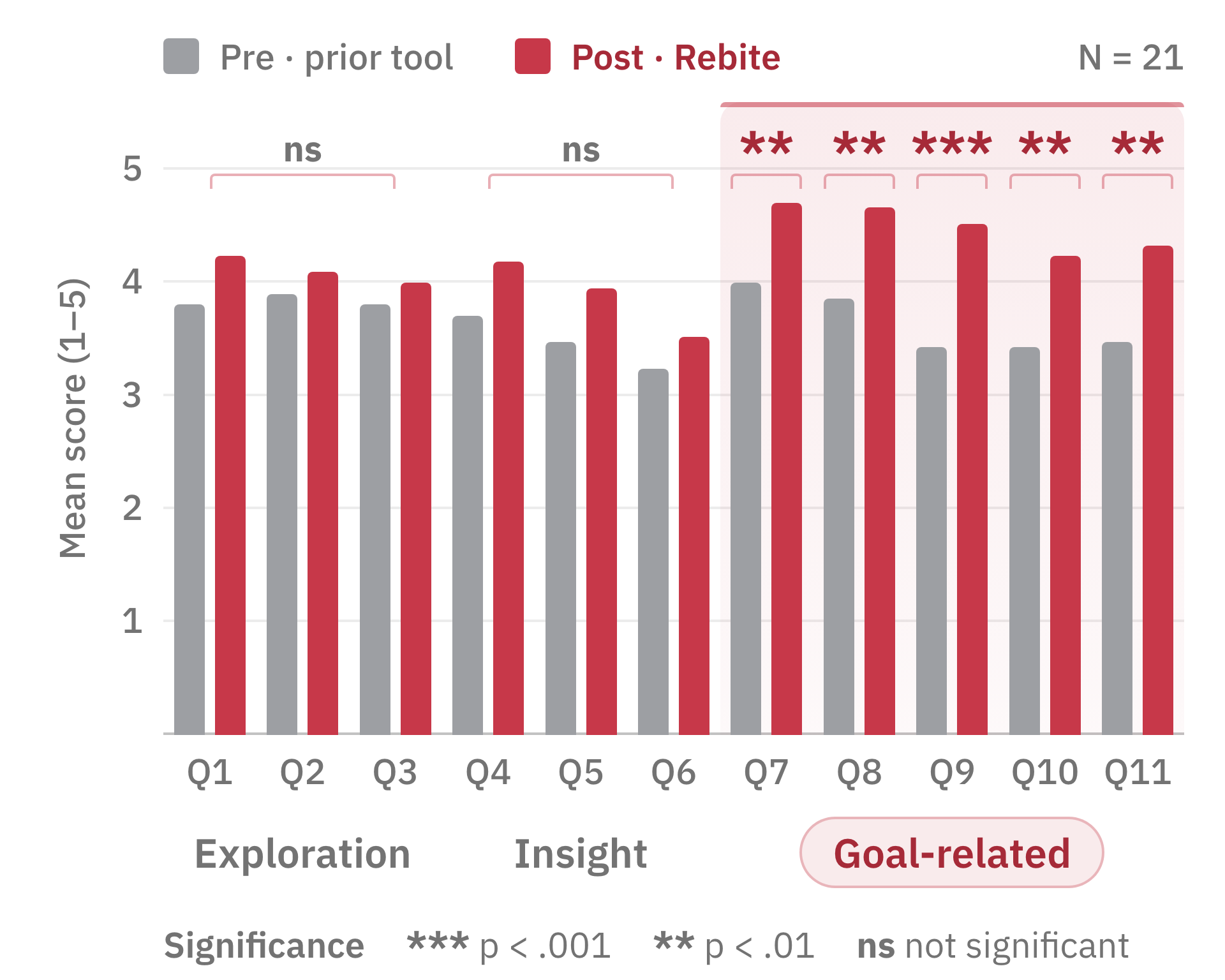}
  \caption{Participants indicated that Rebite assisted with goal-related reflection more than their prior food journaling tools. No significant difference was observed around exploration and insight components of reflection. Significance levels are based on corrected $p$ values.}
  \label{fig:survey}
\end{figure}

Figure~\ref{fig:survey} summarizes the pre--post survey results. The strongest effects appear in the goal-related reflection items (Q7--Q11), where participants scored significantly higher with Rebite than in their recalled ratings of prior tools on all five items ($d_z = 0.75$--$1.10$).

\paragraph{TSRI-derived subscales.}
The Exploration subscale (Q1--Q3) showed a small increase that did not reach significance ($\Delta M=0.27$, $p=.057$, adjusted $p=.078$, $d_z=0.44$). The Insight subscale (Q4--Q6) showed a small increase in the same direction ($\Delta M=0.41$, $p=.039$, $d_z=0.48$), though it did not survive correction for multiple comparisons (adjusted $p=.078$).

\paragraph{Goal-Related Reflection Items (Q7--Q11).}
Participants reported significantly higher awareness of the active dietary goal when using Rebite (\textit{Q7}, $M_{prior}=4.00$, $M_{Rebite}=4.71$, $t(20)=4.18$, $p<.001$, $d_z=0.91$). They also more frequently reported reflecting on whether individual meals fit their current goal (\textit{Q8}, $\Delta M=0.81$, $t(20)=4.25$, $p<.001$, $d_z=0.93$), as well as noticing consistent or inconsistent patterns across meals (\textit{Q9}, $\Delta M=1.10$, $t(20)=5.04$, $p<.001$, $d_z=1.10$). Participants also more often rethought what it means to achieve a dietary goal (\textit{Q10}, $\Delta M=0.81$, $t(20)=3.44$, $p=.003$, $d_z=0.75$). Finally, they reported reflecting more frequently on the goals they were pursuing beyond specific eating behaviors (\textit{Q11}, $\Delta M=0.86$, $t(20)=3.70$, $p=.001$, $d_z=0.81$). The $p$ values reported per item are uncorrected; all five remain significant after Holm--Bonferroni correction across the seven tests (adjusted $p \leq .008$). Participants therefore reported engaging with goals in a wider range of ways with Rebite than with their prior tools.

\subsection{Making Sense of Goal-Directed Feedback~(RQ1)}
\label{sec:goal-translation}

Users arrived with vague goals, such as wanting to ``lose weight'' or ``recover from training''. The system mapped each into a concrete configuration with specific metrics. This mapping offered a foundation for reframing. Because the system determined what to measure, it could later re-determine when goals changed. Three patterns emerged. Participants confirmed what they already understood, expanded their goals, and found mismatches between stated metrics and lived experience.

\begin{figure}[!htbp]
  \centering
  \includegraphics[width=\linewidth]{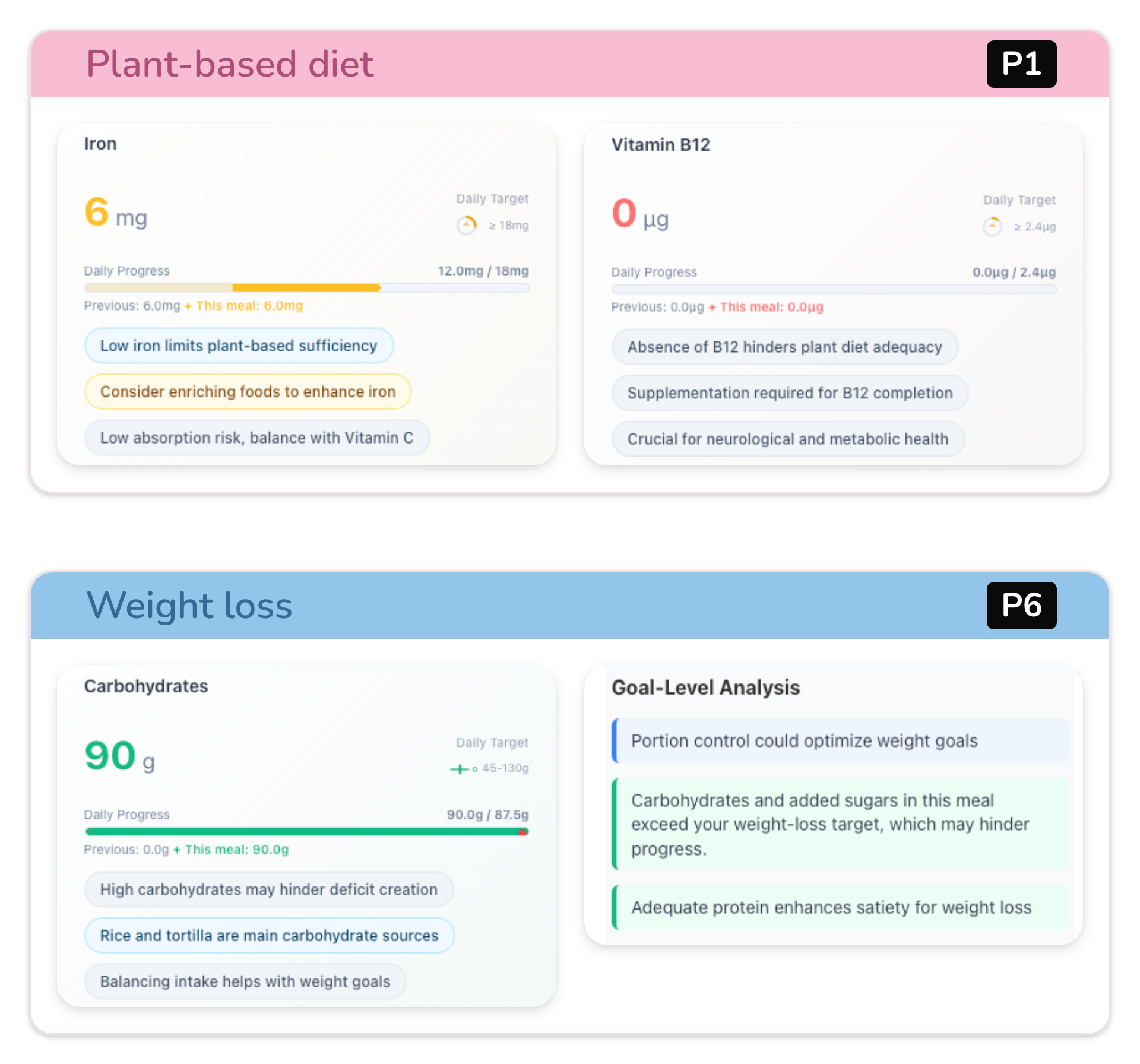}
  \caption{Examples of metrics and analysis surfaced to participants based on their goals. \textbf{P1}'s goal of adhering to a plant-based diet surfaced monitoring iron and B12, while \textbf{P6} monitors carbohydrates to support their weight loss goal.}
  \label{fig:p1_p6}
\end{figure}

\subsubsection{Confirming existing understanding}

When the system surfaced metrics that participants already cared about, it established trust in the translation mechanism (Q7--Q9). P1, a plant-based eater, saw B12 and iron flagged as below target (Figure~\ref{fig:p1_p6}, top). She said, \textit{``B12 and iron are exactly what I always worry about \ldots{} it felt like the system understood me.''} P6 (weight loss) saw carbohydrates highlighted (Figure~\ref{fig:p1_p6}, bottom), and P2 (muscle gain) focused on protein. Participants who trusted the system's initial translation were readier to attend to its re-translation when goals later changed.

\subsubsection{Expanding vague goals into concrete dimensions}

The system also surfaced metrics that participants would not have chosen themselves, making vague goals more concrete (Q10). P21 came in with the goal of ``lose weight'', meaning ``eat fewer calories''. The system added fiber, which she initially accepted out of curiosity, but later found genuinely useful. She said, \textit{``I didn't think much about fiber, but then I noticed it actually kept me full for longer.''} P5's ``anti-inflammatory eating'' became concrete through polyphenols. She explained, \textit{``I had heard the word before, but never linked it to daily food \ldots{} when I go shopping I also check for it now.''} Because the system supplies domain knowledge the user lacks, a new goal supplies \emph{different} knowledge, so reframing produces structurally different feedback from the same data.

However, not all expansions stuck. P4 removed a spice-related metric she found irrelevant, and P1 found fiber unnecessary given her plant-based diet. These cases show the value of letting users adjust the configuration.

\subsubsection{Mismatches as a boundary condition for reframing}

Some goals exceeded the available metric space (Q10--Q11). P19 (recovery nutrition) noted, \textit{``Recovery is also your mood, your training load, your rest \ldots{} eating the right foods doesn't automatically mean I've recovered well.''} P20 (digestive management) felt that FODMAPs, lactose, and fiber missed his experience's complexity. Where the metric set could not represent a goal, reframing under that goal had little to work with.

\subsection{What Happened When Goals Changed~(RQ2)}
\label{sec:reframing-findings}

When participants changed their goal, the system re-mapped it into a new configuration and re-interpreted the same stored meals, with different metrics selected, different assessments assigned, and different interpretations generated. This appeared to make overlaps and conflicts between goals visible through the same structured feedback, prompting participants to negotiate trade-offs, refine priorities, or reject reframing altogether (Q10--Q11).

\subsubsection{Finding overlap between goals}
Some participants found that Rebite helped them see overlap and continuity across goals, showing that the same metrics, foods, or routines could benefit them in more than one way. This gave them a sense that their past efforts were still valuable and reduced the feeling of having to ``start over''.

Some metrics supported multiple goals at once, showing participants that their efforts could serve more than one purpose. For example, P19 explained that when the system highlighted protein intake under his recovery goal, the same \textit{metric card} also showed benefits for blood sugar stability. He reflected, \textit{``When I pushed up protein for recovery, it actually turned out to be good for blood sugar too \ldots{} eat[ing] more meals like this can work for both.''} Before using Rebite, he had focused almost entirely on recovery nutrition, but reframing his logs revealed overlaps across goals.

Others described how the same food could be reinterpreted under different goals. P7, for instance, first saw avocado highlighted under her weight-loss goal as a lower-calorie fat that supported satiety. Later, under her fat-related goal, the \textit{interpretation} praised the same avocado for its unsaturated fat content. She recalled, \textit{``I always thought of avocado as just filling \ldots{} the system showed it as good for fat quality too.''} In some cases, feedback layered affirmations with constructive contrasts. P2, focused on muscle gain, saw her protein shake affirmed as ``supports muscle gain,'' but when she later switched to her balanced-eating goal, the same shake was flagged as ``low diversity.'' She reflected, \textit{``I started using protein powders with added vegetable powders \ldots{} it's an easy way to get protein and still make the meal more balanced.''}

Some participants also carried forward old strategies that remained useful under new goals. P4's low-sugar meals, first adopted for weight loss, later \textit{``also improved my inflammation score.''} Reframing also turned past logs into a resource for learning. P10 explained how old logs expanded in meaning when he added new goals, connecting diet to mental clarity as well as muscle gain. He said, \textit{``I always knew protein was for muscle, but I never thought it connected to mental clarity too.''}

\subsubsection{Surfacing conflict between goals}

When some users switched goals and metrics, Rebite surfaced how the same food could be beneficial for one goal while problematic for the other. Seen through the same feedback format, these contradictions led participants to conclude that a meal has no single, stable reading. P12 described this when she logged salmon. Under \textit{Lose Weight} it was praised for protein, while under \textit{Blood Lipids \& BP} it highlighted omega-3 but also sodium concerns. She reflected, \textit{``I realized the healthy food can be both good and bad.''}

Rebite also surfaced conflicts across daily priorities. P19 focused on endurance cycling, and described how carbohydrates appeared in opposite ways. After logging a pasta dinner post-ride, the \textit{interpretation} under his \textit{Recovery} goal explained, ``Carbohydrates in this meal help restore glycogen after endurance training.'' But under \textit{Blood Sugar Control}, the feedback warned, ``Large carb intake may cause unstable blood sugar.'' He reflected, \textit{``After training I needed carbs, but for blood sugar it flagged the same carbs as a problem.''}

\subsubsection{Revising and refining goals}

As previously described, Rebite made tensions between goals visible by showing different metrics and interpretations under different goals. Instead of treating meals as simply ``good'' or ``bad,'' participants used these conflicts as opportunities to reflect at different levels. They sometimes made small compromises, changed priorities depending on context, reordered goal hierarchies, and redefined what health meant to them.

Some participants adjusted specific choices without abandoning either goal. P17 talked about red bean soup. It felt like a light dessert that worked fine for her weight-loss plan, but it carried enough sugar to concern her blood sugar goal. \textit{``[Rebite]'s tag shows I should reduce the portion \ldots{} Red bean soup is always really low in calories, [it helps me] control my appetite. I just had half a bowl after dinner and put the rest in the fridge for the next day.''} Here, P17 found a middle ground so she could look after both weight and blood sugar.

Other participants refined their goals by switching priorities based on the situation. Facing the carbohydrate conflict described above, P19 created a rule and explained, \textit{``After training I need carbs, but if I'm not training I don't want them to spike my glucose. So I just switch depending on the day.''} Similarly, P13 acknowledged limits in pursuing everything at once. He said, \textit{``I had to accept that I can't maximize everything. Sometimes I lean one [goal], sometimes the other. \ldots{} Right now, [my] weight matters more to me.''} These reflections illustrate how surfaced conflicts may prompt participants to explicitly negotiate trade-offs and refine priorities depending on daily context.

In some cases, Rebite prompted participants to reconsider the hierarchy of their goals over time. P12 reflected, \textit{``I thought weight loss was important, but compared to long-term heart health it's actually less important. I know I can't maximize every goal at once. For me, long-term heart health comes first, then weight loss, and finally anti-inflammation.''} For her, the value of the system was in making conflicts visible without resolving them, enabling her to articulate a new order of priorities. For others, reframing reshaped what health meant in practice. P17 reflected, \textit{``Back then I only cared about calories, if the number was low, I felt good. But looking back, I saw I was always low on protein and fiber. That made me realize eating less wasn't the same as eating well.''}

\subsubsection{Resetting after goal changes}
When participants changed their goals in Rebite, the system resurfaced and reinterpreted past logs. In interviews, some participants said they would rather reset, discarding old logs and starting fresh with their new goal, revealing a tension with the reframing premise that past data should always be reframed. For some, a goal switch created a clear break with the past.

For these participants, old logs felt tied to a finished phase and no longer useful for current decisions. P13 explained, \textit{``If my goal has changed, I don't care what I ate in the past, it's irrelevant now.''} Yet, some showed mixed feelings. P8 wanted a fresh start but was still curious, \textit{``When I set a new goal, I want a fresh start. But I'll eat my favorite food anyway, it's my comfort food. I still want to see how the system explains it under the new goal.''}

Another way participants explained this was as a kind of fresh start ritual, a symbolic act that marked a boundary between ``former me'' and ``current me.'' P3 said switching goals was \textit{``like opening a new one \ldots{} [like] I'm the current me that's really trying to lose weight \ldots{} [it makes me feel] more committed.''} At the same time, participants also described resetting as a way to protect their emotions and motivation. \rev{Part of what made resetting appealing was that reinterpretation could feel like retroactive judgment, where choices that had been reasonable under a previous goal were now recast against the new goal's standard. Some experienced this as a verdict on their past behavior.} P13 explained, \textit{``It felt like something I did before was being judged as wrong \ldots{} like the system was telling me I hadn't done it right.''}

\section{Discussion}
Our study suggests that goal-directed framing can support reflection through a two-step process. The system's goal-to-metric mapping established a concrete, shared language between user and system, which then enabled reframing, where re-interpreting past data under a new goal surfaced overlaps, conflicts, and trade-offs across goals. We discuss how mapping enables reflection, what reframing offers and costs as an interaction, and what it means for data to ``live many lives.''

\subsection{Enabling Reframing: How Goal-to-Metric Mapping Lays the Foundation}

People often start with high-level goals and need help turning them into specific, trackable metrics~\cite{ekhtiar2023goal}. The mapping~(\textbf{D1}) supported exactly this, translating a vague goal into a shared language that the system uses to interpret meals and participants use to understand what their goal means in practice. That shared understanding enabled reframing to support reflection. Because participants already grasped their old goal in concrete terms, when the system re-interpreted the same meals under a new goal, they could compare the two readings and see where goals overlapped or conflicted. This enabled them to read the change as a shift between goals they understood, not an arbitrary system revision.

However, participants noted that this understanding was only as broad as the knowledge base behind the mapping. Some participants emphasized that nutrition-only mappings missed factors like training load or stress, which limited what reframing could surface. Future systems could extend the mapping to incorporate signals beyond nutrition~\cite{li2024vital,nepal2024mindsape}.

Mapping and reframing also reposition the role of the goal. Because Rebite can re-map whenever the goal changes, users do not have to fix their goal in advance~\cite{kim2017omni,sefidgar2024migraine}, and they can revise it in response to what reframing reveals instead of abandoning tracking when it no longer fits~\cite{epstein2015lived,clawson2015no}. In our study, participants negotiated, reordered, and sometimes redefined their goals across goal switches~(\S\ref{sec:reframing-findings}). This suggests that goals can become \emph{outcomes of reflection} instead of fixed preconditions, and that past data stays available to revisit and reflect on as goals change.

\subsection{Designing Reframing: Benefits, Risks, and Implications}

LLMs make it technically feasible to re-interpret unstructured personal data under different goals at scale, but feasibility alone does not determine what the interaction should look like. Our deployment points to three implications.

\subsubsection{Reframing surfaces cross-goal structure that users cannot see on their own}

Reframing surfaced the complexity of pursuing multiple, sometimes conflicting goals. Overlaps reduced the sense of ``starting over,'' while conflicts pushed participants to negotiate trade-offs. Because LLMs can re-assess unstructured data under different goals, these cross-goal relationships can become visible without requiring users to hold multiple framings in memory. This suggests that LLM-powered reframing may extend beyond accommodating new goals, potentially making the \emph{relationships between goals} an explicit object of reflection.

\rev{In our study of Rebite, these cross-goal relationships surfaced sequentially, where users saw them by reframing under each goal in turn, never side by side. This fits how PI is often used in everyday life, favoring brief single-goal glances over deliberate multi-goal comparison~\cite{blascheck2021glanceable}. Still, moments like weighing trade-offs across goals might benefit from a direct cross-goal view. Designing a tool that provides this sort of view in everyday reflection, without burdening it, could be valuable future work.}

\subsubsection{Reframing should support negotiating trade-offs between goals}

Beyond observing cross-goal relationships, participants actively negotiated trade-offs between their goals. They adjusted portions and reordered which goal took priority, sometimes introducing context-dependent rules such as prioritizing recovery on training days and blood sugar on rest days. This negotiation ran on concrete, explained conflict. The four-layer feedback let participants see \emph{which} metric conflicted, \emph{why}, and \emph{which food} caused it, and seeing the same meal assessed under both goals put the trade-off directly in front of them. These natural-language interpretations gave richer anchors than a numeric dashboard alone.

Negotiation therefore depends on the conflict staying visible. Blending multiple goals into a single score would settle the disagreement for participants and remove the very material they reasoned with~(\S\ref{sec:reframing-findings}). Systems should instead keep the conflict intact and help people act on it themselves, for instance by letting them record the context-dependent rules they arrive at, without resolving the trade-off for them.

\subsubsection{Users need control over whether and how much to reframe}

Reframing also caused discomfort for some. Some participants preferred to reset entirely, treating previous logs as a closed chapter (e.g., happy abandonment~\cite{clawson2015no,epstein2016abandon}). Others felt uneasy when feedback recast previously ``healthy'' foods as problematic. Because re-interpreting all stored data at once is technically cheap, a single goal change can deliver far more re-assessment than the user asked to see.

\rev{This discomfort shows that reframing does not merely re-describe past behavior. It re-evaluates that behavior against a \emph{new} standard. The same content points two ways. As advice about the next meal it feels actionable, but applied to meals already eaten it can read as a verdict, and some participants experienced it as being judged for past choices~(\S\ref{sec:reframing-findings}). For some, that is why they preferred to reset to escape feelings of judgment.}

\rev{Our instantiation of goal-directed reframing in Rebite left users with a largely binary choice of accepting the new assessment or resetting and walking away. Neither option lets a user agree with part of a re-assessment while pushing back on another, or contest a specific verdict without abandoning the broader reframing. A first step towards resolving this tension is giving users control over the \emph{scope} of reframing by indicating which data to re-interpret, at what granularity, and whether to engage at all. Beyond scope, users would also benefit from ways to negotiate with the re-assessment itself. Prior HCI work has made system output adjustable instead of take-it-or-leave-it~\cite{lee2022sleepguru} and let users revise their own input after the fact~\cite{le2025recall,hoefer2025telltime}. Translated to reframing, a re-assessment could allow users to accept some reinterpretations, defer or reject others, or annotate why a past choice was reasonable under its original goal, so they keep what reframing offers while staying in control of their own data.}

\subsection{Generalizing Reframing: When Data Lives Many Lives}

\subsubsection{The same data can mean many things}

Goal-directed framing broadens what PI systems can do with already-collected data. By keeping raw data and deferring interpretation, the same meal can be framed as recovery fuel, a weight-management risk, or a contribution to dietary diversity without new data entry. Prior critiques have shown how tracking tools can reproduce narrow narratives, such as equating health with thinness~\cite{10.1145/3613904.3642600,Eikey2017ed}. Keeping data open to reinterpretation may help resist these rigidities by treating health as situational and changeable.

However, LLMs also carry risks of amplifying harmful narratives. Biased outputs may reinforce restrictive dieting or pro-eating disorder tendencies, tied to cultural biases embedded in the models themselves. Anchoring metric selection to published guidance constrains this only partly. Our framework is authored, not clinically validated, and fixing the metric set does not fix the language the model uses about it. Whether goal plurality widens the range of values a system can express, or only gives a biased model more surfaces to speak from, is an open question. Reinterpreting the same data under different goals also shows that no assessment is neutral. The system's interpretation, the user's own understanding of the meal, and the values built into the active goal can diverge, and reframing surfaces those differences for the user to weigh.

\subsubsection{The cost of deferring interpretation}

\rev{This ``capture now, interpret later'' stance is double-edged. Decoupling capture from any single goal lowers the barrier to logging and keeps data useful when priorities change, which is much of what makes reframing possible in the first place. Yet it sits uneasily with a long-standing view in personal informatics that \emph{why} people record shapes \emph{what} they record~\cite{epstein2015lived}. If the ``why'' is deferred indefinitely, the motivation that sustains careful logging may erode, and capture may drift toward data hoarding. How to keep open-ended logging worthwhile is a question future work should examine. Automated and semi-automated collection through wearables and passive sensing makes preserving rich data increasingly feasible, which only sharpens this question, as preserving reusable data may be necessary but not sufficient. Future systems may therefore need to support goal \emph{discovery} and refinement~\cite{ekhtiar2025goal}, letting users explore data and goals together so that each decision improves the other over time.}

\subsubsection{Beyond food journaling}

Unclear and changing goals are not unique to food journaling~\cite{ekhtiar2025goal}. The approach may generalize to domains that satisfy three conditions. Raw data supports multiple interpretations (e.g., sleep sensor data encodes latency, efficiency, stages, and HRV), a structured knowledge base can map goals to relevant dimensions (e.g., clinical guidelines for sleep hygiene vs. athletic recovery), and factual estimates can be separated from goal-directed assessments. Prior work has shown that personalized systems can adapt to individual users in domains like sleep~\cite{daskalova2016sleepcoacher} and conversational goal tracking~\cite{ongoal2025}. Extending to a new domain would likely require building a domain-specific mapping layer, though the rest of the pipeline could be reused.

\section{Conclusion}
We introduce goal-directed framing, in which the goal acts as an active lens over preserved data, and realize it in Rebite. Reframing past meals under a new goal surfaced overlaps and conflicts that prompted users to negotiate trade-offs and refine priorities, though some preferred to start fresh. The approach could extend to domains such as exercise, sleep, and mental health.

\begin{acks}
This work was partially supported by the National Science Foundation under award IIS-2237389. Generative AI tools were used for grammar and sentence editing, advice on shortening, and some development assistance.
\end{acks}

\bibliographystyle{ACM-Reference-Format}
\bibliography{sample-base}

\end{document}